\documentclass[journal]{IEEEtran}

\usepackage[OT1]{fontenc} 
\usepackage[numbers,sort&compress]{natbib}
\usepackage[cmex10]{amsmath}
\usepackage{amssymb}
\usepackage{bm}
\usepackage{braket}
\usepackage{graphicx}
\usepackage{xcolor}
\usepackage{subfigure}
\usepackage{tabularx}
\usepackage{arydshln}
\usepackage{mathtools}
\usepackage{multirow}
\usepackage{algorithm,algorithmic}
\usepackage{url}
\usepackage{listings}
\usepackage{comment}
\usepackage{CJKutf8}
\usepackage[normalem]{ulem}

\usepackage{hyperref}
\usepackage[absolute,overlay]{textpos}
\makeatletter
\def\ps@IEEEtitlepagestyle{%
  \def\@oddfoot{\mycopyrightnotice}%
  \def\@oddhead{\hbox{}\@IEEEheaderstyle\leftmark\hfil\thepage}\relax
  \def\@evenhead{\@IEEEheaderstyle\thepage\hfil\leftmark\hbox{}}\relax
  \def\@evenfoot{}%
}
\def\mycopyrightnotice{%
  \begin{minipage}{\textwidth}
  \centering \scriptsize
\textcopyright 2025 IEEE. Personal use of this material is permitted.
  Permission from IEEE must be obtained for all other uses, in any current or future
  media, including reprinting/republishing this material for advertising or promotional
  purposes, creating new collective works, for resale or redistribution to servers or
  lists, or reuse of any copyrighted component of this work in other works.
  DOI: \href{https://doi.org/10.1109/MCOM.001.2500390}{10.1109/MCOM.001.2500390}
  \end{minipage}
}
\makeatother
\begin{document}
\title{Quantum-Accelerated Wireless Communications: Concepts, Connections, and Implications}

\author{Naoki~Ishikawa,~\IEEEmembership{Senior~Member,~IEEE}, Giuseppe~Thadeu~Freitas~de~Abreu,~\IEEEmembership{Senior~Member,~IEEE},\\ 
Petar~Popovski,~\IEEEmembership{Fellow,~IEEE}, and Robert~W.~Heath~Jr.,~\IEEEmembership{Fellow,~IEEE}.\thanks{N.~Ishikawa (corresponding author) is with the Faculty of Engineering, Yokohama National University, Japan.
G.~T.~F.~de~Abreu is with the School of Computer Science and Engineering, Constructor University, Germany.
P.~Popovski is with the Department of Electronic Systems, Aalborg University, Denmark.
R.~W.~Heath is with the Department of Electrical and Computer Engineering, the University of California, San Diego, USA.}}

%\markboth{\today}
%{Shell \MakeLowercase{\textit{et al.}}: Bare Demo of IEEEtran.cls for Journals}
\maketitle

%ARXIV
\TPshowboxesfalse
\begin{textblock*}{\textwidth}(45pt,10pt)
\footnotesize
\centering
Accepted for publication in IEEE Communications Magazine. This is the author's version which has not been fully edited and content may change prior to final publication. Citation information: DOI 10.1109/MCOM.001.2500390
\end{textblock*}
%ARXIV

\begin{abstract}
Quantum computing is poised to redefine the algorithmic foundations of communication systems. While quantum superposition and entanglement enable quadratic or exponential speedups for specific problems, identifying use cases where these advantages yield engineering benefits is still nontrivial. This article presents the fundamentals of quantum computing in a style familiar to the communications society, outlining the current limits of fault-tolerant quantum computing and clarifying a mathematical harmony between quantum and wireless systems, which makes the topic more enticing to wireless researchers. Based on a systematic review of pioneering and state-of-the-art studies indicating speedup opportunities, we distill common design trends for the research and development of quantum-accelerated communication systems and highlight lessons learned. The key insight is that quantum algorithms, including their gate-level realizations, can benefit from the design intuition applied in communication engineering. This article aims to catalyze interdisciplinary research at the frontier of quantum information processing and future communication systems.
\end{abstract}

\IEEEpeerreviewmaketitle

\section{Introduction}
\label{sec:intro}
It is known that quantum computing (QC) can dramatically accelerate the simulation of quantum dynamics and chemical processes, and perform efficient prime factorization. But where might QC help communications beyond cryptography?
Notice that our motivating question differs from that pursued by quantum machine learning approaches relying on strong assumptions of idealized oracles and cost-free quantum data access \cite{tang2022dequantizing}.
In particular, while classical computers are known to struggle with NP-hard problems that often arise in communications systems, and although QC is not known to overcome this limitation, we address whether QC can offer quadratic speedups for unstructured exhaustive search \cite{grover1996fast}, including certain NP-hard problems that admit efficiently implementable oracles \cite{gilliam2021grover}, rather than relying on idealized oracles.

Realizing such advantages at scale requires fault-tolerant quantum computing (FTQC), whose substantial resource demands remain a major barrier.
Although the hardware overhead remains formidable,
recent studies, most notably the finding that 2048-bit RSA integers can be factored in about one week with fewer than one million noisy qubits \cite{gidney2025how}, show that the barrier is steadily shrinking.

The objective of this article is to provide an overview of QC, with a particular focus on its mathematical foundations and potential applications relevant to the communications society.
Some of the mathematical foundations share interesting similarities with wireless communication theory.
A representative quantum algorithm capable of achieving quadratic speedups is then introduced.
From a systematic review, we distill common trends for transplanting quantum and its related techniques into complex problems in communication systems, and provide lessons learned.
Finally, we outline the open challenges that await at the intersection of QC and future communication systems.

\section{Quantum Fundamentals from a Communications Perspective}
\label{sec:basics}

Wireless researchers familiar with linear algebra can quickly grasp the fundamentals of QC by drawing parallels communication-theoretic tools.

\subsection{Qubits, Quantum States, and Gates}
\label{subsec:qubits}

Quantum and classical computing differ fundamentally in how information is represented and manipulated.
In particular, in QC, a quantum bit, \emph{qubit}, serves as the fundamental unit of information. It can exist in a state that is a superposition of two definite states denoted by $\ket{0}$ and $\ket{1}$. Specifically, a quantum state can be expressed as a linear combination
\begin{align}
    a \ket{0} + b \ket{1},
\end{align}
where $\ket{0}$ and $\ket{1}$ are the standard basis (or computational basis) vectors in two-dimensional space, $i.e.$, the first and second columns of the identity matrix.

The two numbers $a$ and $b$ above are complex values, which determine how much the qubit leans toward each of the two possible states.
When measured in the computational basis, the state collapses to $0$ with probability $|a|^2$, or to $1$ with probability $|b|^2$, such that these probabilities must add up to $1$.
In superconducting platforms, control and readout are naturally described in the I/Q plane: the microwave in-phase (I) and quadrature (Q) components shape control pulses and, via homodyne or heterodyne detection, indicate whether the state is in $\ket{0}$ or $\ket{1}$.
Once the measurement is made, the original state is irreversibly lost.

The fact that the qubit is a function of the two continuous complex values $a/b$ may entice a communication engineer to modulate information into $a$ and $b$.
In \textit{amplitude encoding} of QC, classical data are modulated into the probability amplitude $|a|^2$ (or $|b|^2$), much like pulse amplitude modulation in classical communications, in an algebraic sense.
In \textit{phase encoding}, data are modulated into the relative phase between $a$ and $b$, mirroring differential phase-shift keying.
Practical advantages appear when many identically prepared qubits are processed.
Beyond these, there are several other operations that map classical data to quantum states.
Such techniques are often called \textit{quantum embeddings}.

\begin{figure}[tb]
\centering
\includegraphics[width=0.9\linewidth]{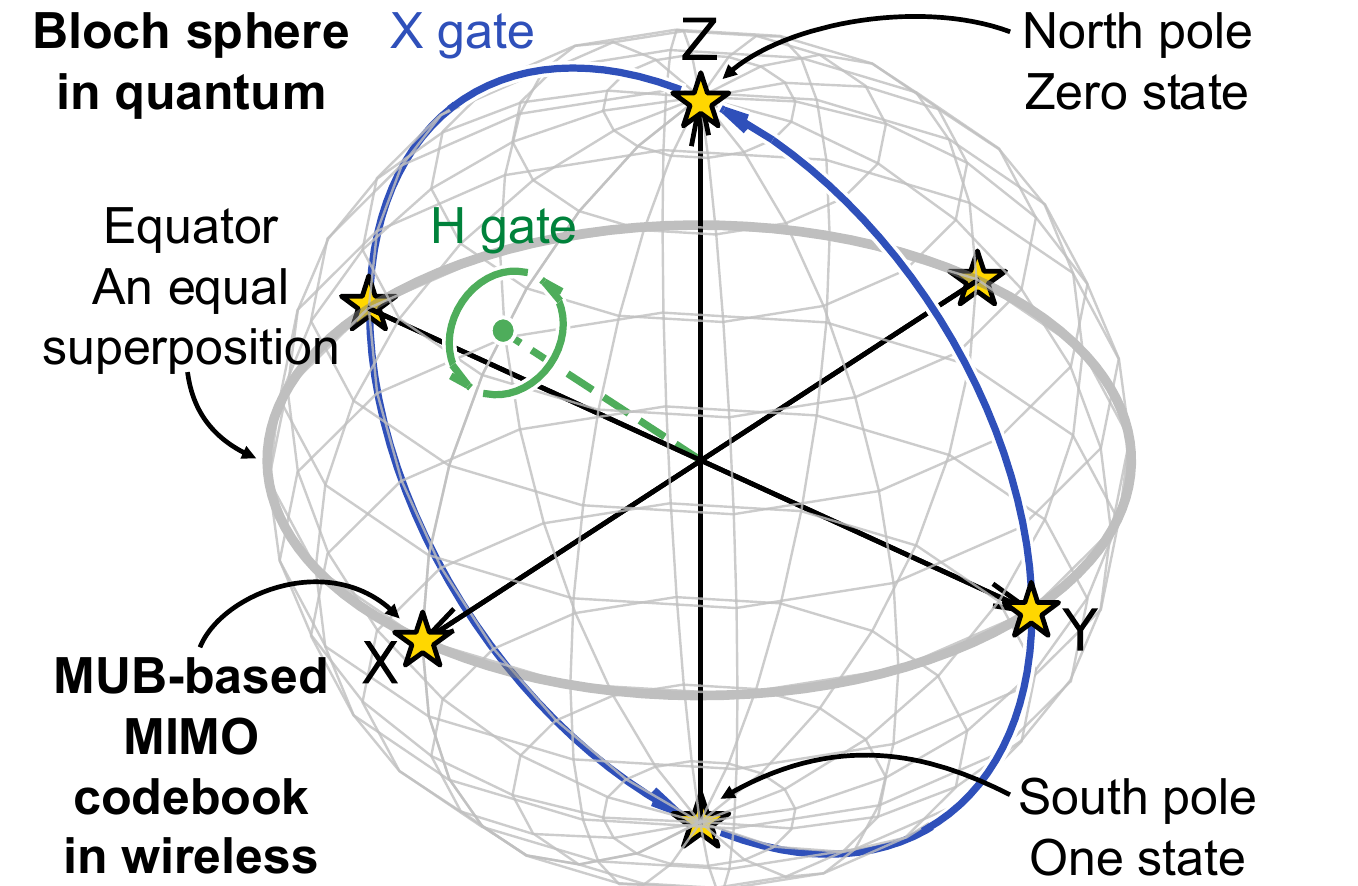}
\caption{The Bloch sphere in quantum computation and a Grassmannian codebook in wireless communications. A single-qubit quantum gate operation can be represented as a rotation of a quantum state about an arbitrary axis of the Bloch sphere. The MIMO codebook here is known as MUB in quantum measurement theory.}
\label{fig:bloch} 
\end{figure}

The natural constraint that probabilities sum to one allows a quantum state to be represented on the surface of a 3D sphere called the Bloch sphere, as shown in Figure 1.
If we think of the Bloch sphere as a globe, the zero state $\ket{0}$ is at the north pole, the one state $\ket{1}$ is at the south pole, and their equal superposition corresponds to any point on the equator.
A quantum computer applies quantum gates to manipulate a qubit, with each gate operation corresponding to the rotation of a quantum state about an arbitrary axis of the Bloch sphere.

\subsection{Quantum Superposition}
\label{subsec:superposition}
Wireless researchers may find the following Hadamard matrix familiar:
\begin{align}
    \mathrm{H} = \frac{1}{\sqrt{2}} \begin{bmatrix}
        +1 & +1 \\
        +1 & -1
    \end{bmatrix}.
\end{align}
An equal superposition can be generated by such a Hadamard matrix, called the H gate.
When the H gate is applied to the state $\ket{0}$, it results in an equal superposition of $\ket{0}$ and $\ket{1}$. 
As illustrated in Figure 1, the H gate corresponds to a 180-degree rotation about an axis that lies at a 45-degree angle between the X and Z axes. 
When applied to $\ket{0}$, it moves the state to a point halfway between $\ket{0}$ and $\ket{1}$, located at the front of the X axis.
Naturally, applying the H gate again to that state brings it back to the zero state.

\begin{figure}[tb]
\centering
\includegraphics[width=1\linewidth]{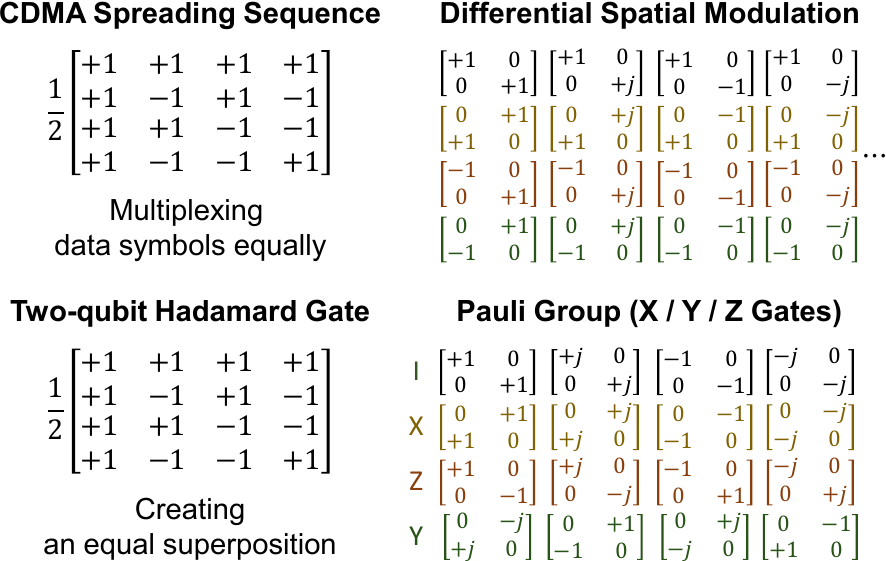}
\caption{Similarities between quantum and wireless systems, such as equal superposition and sparse unitary matrices.}
\label{fig:similar}
\end{figure}

Operations on the Bloch sphere can be also represented using basic linear algebra.
Since the states $\ket{0}$ and $\ket{1}$ correspond to the first and second columns of the identity matrix, respectively, the application of the H gate can be understood by $\mathrm{H} \ket{0}$ and $\mathrm{H} \mathrm{H} \ket{0}$.
For a two-qubit system, such as one initialized at the state $\ket{00}$, the two-qubit Hadamard gate is extended by taking the tensor product of the two Hadamard matrices.
Interestingly, as illustrated in Figure~\ref{fig:similar}, this four-by-four matrix mirrors the spreading sequences used in CDMA with four user terminals, which is quite familiar to wireless researchers.
Note that this similarity lies in the concept of linear superposition, not in the underlying physical mechanisms.

Generally, while in CDMA symbols from $2^n$ users are superposed with equal weighting, in QC an equal superposition of all $2^n$ possible states of a set of $n$ qubits is represented by an $n$-qubit register.
For such an equal superposition, a quantum algorithm repeatedly applies quantum gate operations, selectively enhancing the amplitudes of desired states through constructive interference, while suppressing the undesired ones via destructive interference thus increasing the probability of observing the desired states.
This procedure is similar to a wireless communication technique that uses linear operations to direct signal power spatially and suppress interference in undesired directions, while keeping total transmission power constant.

\subsection{Quantum Entanglement}\label{subsec:entanglement}
Quantum entanglement is a phenomenon akin to having two magic coins that always match.
For a two-qubit system, such entanglement can be generated by an H gate and a controlled-NOT (CNOT) gate, also known as a controlled-X gate.

To elaborate, consider an X gate which, acting on a single qubit, flips $\ket{0}$ to $\ket{1}$, and $\ket{1}$ to $\ket{0}$.
In turn the CNOT gate operates on a pair of qubits: one designated as the control and the other as the target.
If the control qubit is in the zero state, the gate leaves the target qubit unchanged.
In contrast, if the control qubit is in the one state, the X gate is applied to the target qubit.
If the control qubit of a CNOT gate is in an equal superposition of $\ket{0}$ and $\ket{1}$, and the target qubit is in the zero state, the resultant state is known as a Bell state.\footnote{Note that the term honors physicist John Stewart Bell and is unrelated to Alexander Graham Bell.}
Measuring the state yields outcomes $00$ or $11$ with equal probabilities of $1/2$.
That is, measuring either qubit individually results in random outcomes ($0$ or $1$), but upon measurement of one qubit, the state of the other qubit immediately becomes certain, which is said to be entangled.

Manipulating quantum states with unitary gates has striking parallels with the use of unitary matrices in differential encoding for noncoherent wireless communications.
The X gate corresponds to a simple two-by-two anti-diagonal matrix.
Just as this matrix represents a 180-degree rotation about the X-axis on the Bloch sphere, rotations about the Y and Z axes are implemented by the Y and Z gates, respectively.

As illustrated in Figure~\ref{fig:similar}, these matrices form a closed group under matrix multiplication, which is called the Pauli group.
A similar idea is independently used in wireless communications via sparse-unitary differential spatial modulation \cite{bian2015differential}.
A detailed discussion can be found in \cite{cuvelier2021quantum}. Specifically, a new class of space-time block codes is designed by the X, Y, and Z gates, illustrating the deep interrelation between quantum and wireless systems.

\section{A Hidden Mathematical Harmony in Quantum and Wireless Systems}\label{sec:harmony}
As described in Section~\ref{sec:basics}, single-qubit operations correspond to rotations on the Bloch sphere shown in Figure~\ref{fig:bloch}.
In turn, multi-qubit operations can be viewed as movements on a Grassmann manifold, which underlies the mathematical similarities between quantum and wireless systems.

In wireless systems, the problem of designing an optimal MIMO precoding codebook under Rayleigh fading can be formulated as an optimization problem over a Grassmann manifold \cite{love2003grassmannian}.
For example, when transmitting a single data stream using multiple antennas, unit-norm precoding vectors are used to maintain constant average transmit power.
Precoding vectors differing only by a global phase achieve the same average mutual information, forming an equivalence class.

The set of all such equivalence classes can be interpreted as a quotient space known as the complex Grassmann manifold.
For a given codebook size, the optimal codebook design is a packing problem that involves maximizing the minimum distance among codewords uniformly distributed on the Grassmann manifold.
A commonly used distance metric in this context is the chordal distance, the length of the chord connecting two codewords (or subspaces), especially when maximizing average mutual information under Rayleigh fading.

Interestingly, the geometric structure of Grassmann manifolds in wireless communications also appears in representing (pure) quantum states.
In particular, a global phase rotation applied to a quantum state vector, $i.e.$, the linear combination of $\ket{0}$ and $\ket{1}$, does not affect the outcome probabilities of quantum measurements.
This means that all quantum state vectors differing only by a global phase represent the same physical state.
As in the wireless case, each quantum state can be represented as a unit vector modulo a global phase, corresponding to a quotient space.
Therefore, single- or multi-qubit pure states are naturally represented within a complex Grassmann manifold structure.

This algebraic similarity allows a Grassmannian MIMO precoding codebook, or a Grassmannian constellation designed for noncoherent communications, both composed of two-by-one vectors, to be visualized on the Bloch sphere used in QC.
An example codebook is shown in Figure~\ref{fig:bloch}, which consists of six two-by-one vectors obtained by maximizing the minimum chordal distance between points.
This uniform placement improves performance in wireless communications.
Surprisingly, the codebook in Figure~\ref{fig:bloch} is identical to the mutually unbiased bases (MUB) used in quantum measurement theory.

A key metric in QC is fidelity, which quantifies the similarity between two quantum states, and can be characterized using the chordal distance on the Grassmann manifold.
Fidelity is also used as a performance metric for quantum devices, indicating how accurately the implemented state reproduces the intended target state.
Given two pure quantum states, each represented as a point on the complex Grassmann manifold, the fidelity $F$ is calculated as the squared magnitude of the inner product between them.
By contrast, the chordal distance is given by 
\begin{equation}
d = \sqrt{1 - F},
\end{equation}
highlighting a direct mathematical connection.
Note that these discussions do not extend to mixed quantum states.

\section{Grover-based Algorithms for Quantum Optimization}\label{sec:grover}
A variety of approaches have been proposed to use quantum mechanics and its related technologies to tackle NP-hard problems.
Special-purpose physical machines for such problems includes the quantum annealer (QA) developed by D-Wave and the coherent Ising machine (CIM) developed by NTT and Stanford University.
Both QA and CIM obtain approximate solutions to quadratic unconstrained binary optimization (QUBO) problems, with QA already available on the public cloud, while CIM is not commercially available yet.
Although neither offers proven speedup or optimality guarantees, they can produce high-quality approximate solutions in a sufficiently short time \cite{hamerly2019experimental}, making them attractive from an engineering perspective.

As for algorithms that run on general gate-based quantum computers, well-known approaches are the quantum approximate optimization algorithm (QAOA) and the Grover-based approach.
The QAOA approach is designed with today's noisy intermediate-scale quantum (NISQ) computers and therefore has the advantage of being supported by many experimental demonstrations.
However, the principle of QAOA is rooted in the quantum adiabatic theorem that also underpins QA, such that the approach shares the same limitation of lacking speedup guarantees, although it can still produce high-quality approximate solutions in some cases.

The Grover-based approach can be executed on today's NISQ devices at small scale, while large-scale instances will ultimately require FTQC.
Grover's algorithm offers a provably optimal quadratic speedup in query complexity for unstructured search.
Through its generalized and extended algorithms, it serves as a versatile sub-routine that accelerates many search-type and combinatorial optimization problems.
The failure probability can be driven down exponentially, and the optimal solution can be found with very high probability.

\begin{figure}[tb]
\centering
\includegraphics[width=1\linewidth]{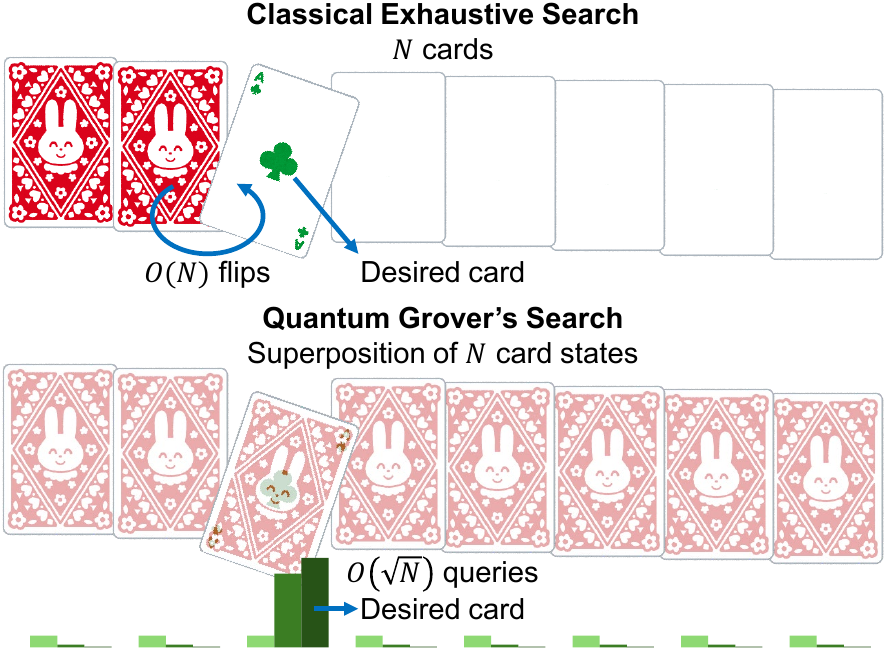}
\caption{Card-flipping analogy illustrating Grover's quadratic speedup. A classical exhaustive search requires, in the worst case, $N$ flips to reveal the desired card, whereas Grover's algorithm finds it with about $\sqrt{N}$ queries.}
\label{fig:card}
\end{figure}

The underlying mechanism of Grover's search may seem abstract or even difficult to grasp at first.
To build some intuition, let us explain its principle via the following game.

Imagine that there is a deck of $N$ cards, only one of which is desired, as illustrated in Figure~\ref{fig:card}.
Each move of a classical player consists of flipping a single card.
With a classical exhaustive search, the classical player is therefore expected to find the desired card after flipping, on average, $N/2$ cards.
In the worst case, all $N$ cards must be flipped, resulting in a worst-case time complexity of $O(N)$.

In contrast, each move of a quantum player consists of peeling off a layer of the back of all cards simultaneously, making them slightly transparent.
After performing such operation repeatedly, approximately $\sqrt{N}$ times, the quantum player can identify the exact position of the desired card in the deck.
The operation here paraphrased by ``peeling the back of all cards simultaneously'' corresponds to performing a query to an oracle, and the total number of such oracle queries determines the query complexity of the overall procedure.
Now, if we assume that flipping a card has the same cost as, or within a constant factor of, making a query to the oracle, the quantum player makes $O(\sqrt{N})$ moves, against $O(N)$ moves of the classical player.
This improvement is commonly referred to as a quadratic speedup.

\begin{figure*}[tb]
\centering
\includegraphics[width=1\linewidth]{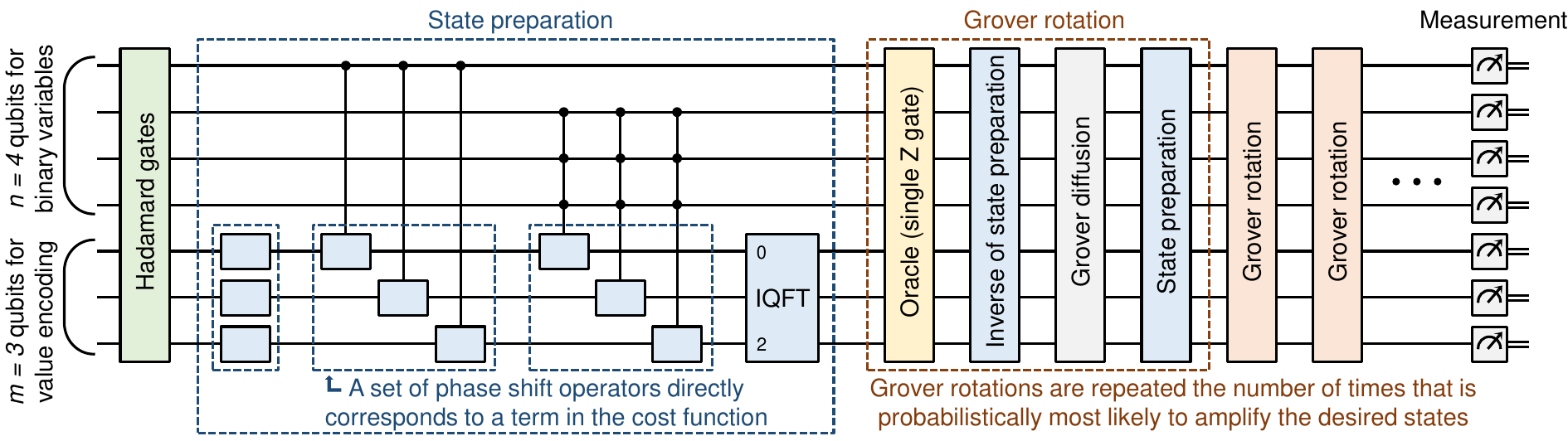}
\caption{An overview of the quantum circuit for Grover adaptive search, where $n=4$ qubits hold candidate solutions and $m=3$ qubits encode their costs as phases. An inverse quantum Fourier transform (IQFT) turns those phases into amplitudes, then repeated oracle-flip and Grover diffusion operations amplify the desired state, which is read out by measurement.}
\label{fig:gas}
\end{figure*}

From a geometric viewpoint, Grover's algorithm follows the shortest path (geodesic) from an initial quantum state to a solution state on a Grassmann manifold, and dividing the distance between the two states by the per-query advance yields approximately $\pi\sqrt{N}/4$ \cite{miyake2001geometric}.

The execution of Grover's algorithm and its adaptive extension for global minimization, Grover adaptive search (GAS), require the construction of an appropriate oracle, which is a quantum circuit that identifies whether a given quantum state satisfies the desired condition or not.
Concrete and efficient methods for constructing quantum circuits for oracles corresponding to specific problems have been proposed.
A breakthrough in this area is related to a binary polynomial optimization problem \cite{gilliam2021grover}, which seeks to minimize a cost function of $n$ binary variables.
Due to their NP-hard nature, in many binary optimization problems, all $2^{n}$ cost values need to be calculated to obtain the exact solution.

The quantum circuit of a GAS algorithm \cite{gilliam2021grover} is illustrated in Figure~\ref{fig:gas}.
The circuit solves a binary polynomial optimization on $n+m$ qubits where the first $n$ qubits hold the binary variables, while the remaining $m$ store costs in two's-complement form, with the sign in the most-significant qubit,
enabling the identification of desired states, $i.e.$, those with negative costs.
Accordingly, the oracle is constructed by a single Z gate, which is applied to the most-significant qubit, flipping the phase of exactly desired negative states.
Each cost-function term maps to a specific gate set, and it can handle QUBO and higher-order unconstrained binary optimization (HUBO) problems.
The original proposal assumed integer coefficients, but later work \cite{norimoto2023quantum} extended the mapping to arbitrary real values.

The GAS algorithm begins with a uniform superposition created by the H gates, and all $2^{n}$ costs are calculated in parallel using the conditional behavior of CNOT gates and specific gates.
This is called \textit{quantum parallelism} and is often misunderstood that it alone yields a quantum advantage.
It does not, because measurement returns a single outcome whose probability is only $1/2^{n}$.
That is, all $2^{n}$ costs are evaluated simultaneously, but only one of them can be observed.
The algorithm can amplify the probability of desired states in $O(\sqrt{2^{n}})$ operations.
GAS applies the boost adaptively, shifting a constant term so that fewer states remain negative until only the global minimum survives.
The query complexity is still exponential, yet the quadratic speedup effectively halves the problem size, a gain that matters on large instances.

\section{Case Studies and Lessons Learned}
\label{sec:case}

\begin{figure*}[tb]
\centering
\includegraphics[width=1\linewidth]{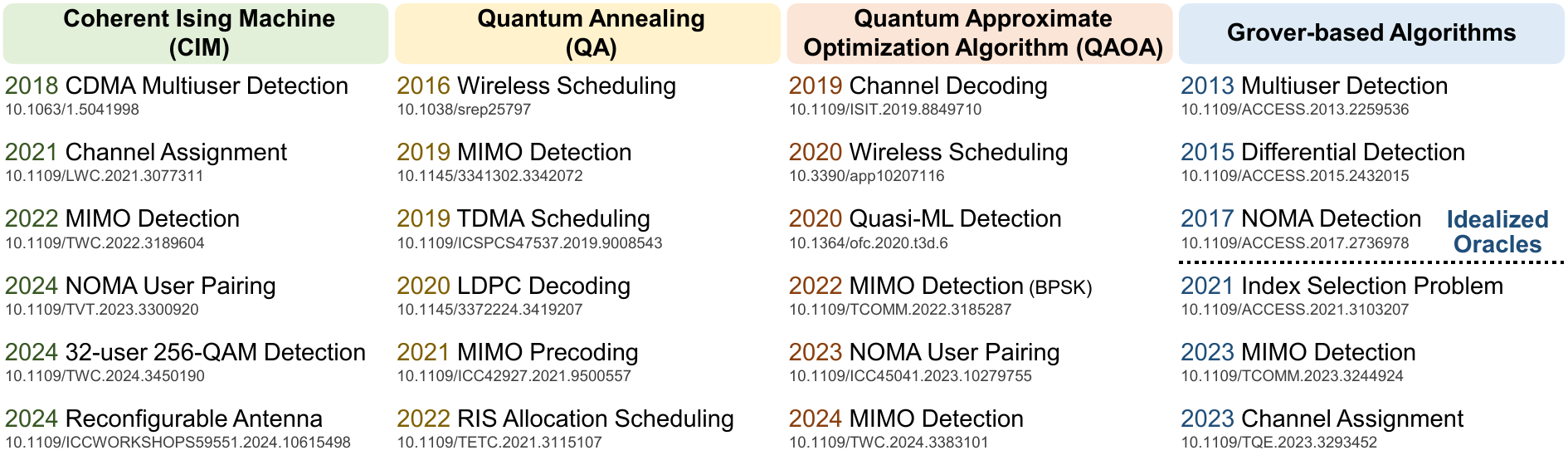}
\caption{Timeline of early pioneering studies on the application of CIM, QA, QAOA, and Grover-based algorithms to communication systems.}
\label{fig:timeline}
\vspace{-2ex}
\end{figure*}

The application of QC to communication problems dates back to the early 2010s \cite{botsinis2019quantum}, led by pioneering efforts from a research group at the University of Southampton.
More than a decade after this series of studies, it has become possible to construct specific oracles in an efficient manner, as described in the previous section.
Figure~\ref{fig:timeline} presents a timeline categorizing such early pioneering studies by CIM, QA, QAOA, and Grover-based algorithms including GAS.
It shows the evolution of quantum-related studies across the physical, data link, and network layers, and encourages co-design of classical and QC for future communication systems.
From these studies, three common trends can be identified.

First, most studies target NP-hard discrete optimization problems, because such problems require, in the worst case, exponential time in classical computing, while problems solved in polynomial-time via classical algorithms generally do not warrant a quantum advantage.
However, for continuous problems, quantum algorithms such as the Harrow-Hassidim-Lloyd algorithm and quantum gradient descent can still offer provable speedups under certain assumptions.

Second, most problems are formulated as QUBO or HUBO problems.
In particular, CIM and QA supports QUBO problems, whereas QAOA and GAS can also support HUBO problems that contain higher-order terms.
Although formulations using binary variables and spin variables are mathematically equivalent, the number of terms in the cost function can differ.
Even when the obtained formulation seems identical to a well-known combinatorial optimization problem, communication-specific structure such as channel randomness, interference, and network dynamics are often exploited.

Third, benchmarking is delicate.
QA, CIM, and QAOA can be evaluated by wall-clock time and approximation ratio, since these approaches are designed to obtain approximate solutions running on actual hardware.
Strong classical baselines such as efficient approximation algorithms, semidefinite relaxations, meta-heuristics, and machine learning already deliver fast, high-quality approximations for small- to medium-scale NP-hard problems.
Since a Grover-based algorithm can be regarded a quantum exhaustive search that achieves quadratic speedup while still finding the optimal solution, it is one of the few cases where a direct comparison with classical exhaustive search is both fair and informative.

From the studies summarized in Figure~5, we classify them into two categories and present several cases from each.

\subsection{Case Study 1: Detection / Decoding}\label{subsec:case1}

\paragraph{Problem and challenges}
Maximum-likelihood detection and channel decoding in wireless systems are both computationally hard due to the exponential growth of the search space with the transmission bits.
While practical systems adopt linear or low-complexity methods to reduce latency, these come at the cost of reduced accuracy.
QC offers potential speedup, but requires careful problem formulation that avoids search space expansion.

\paragraph{Key contributions}
A pioneering study from Princeton University \cite{kim2019leveraging} formulated MIMO detection as a binary optimization problem and solved it with QA, attaining near-optimal performance that exceeded a conventional linear detector.
In the Grover-based approaches, an explicit oracle for MIMO detection was first constructed in \cite{norimoto2023quantum}.
The internal parameters of GAS were determined according to the cost-function distribution analysis and a classical approximate solution.
For polar decoding \cite{fujiwara2024quantum}, the initial quantum state was prepared as a uniform superposition of only valid codewords, avoiding unnecessary expansion of the search space and achieving a pure quadratic speedup.

\begin{figure}[tb]
\centering
\includegraphics[width=1.0\linewidth]{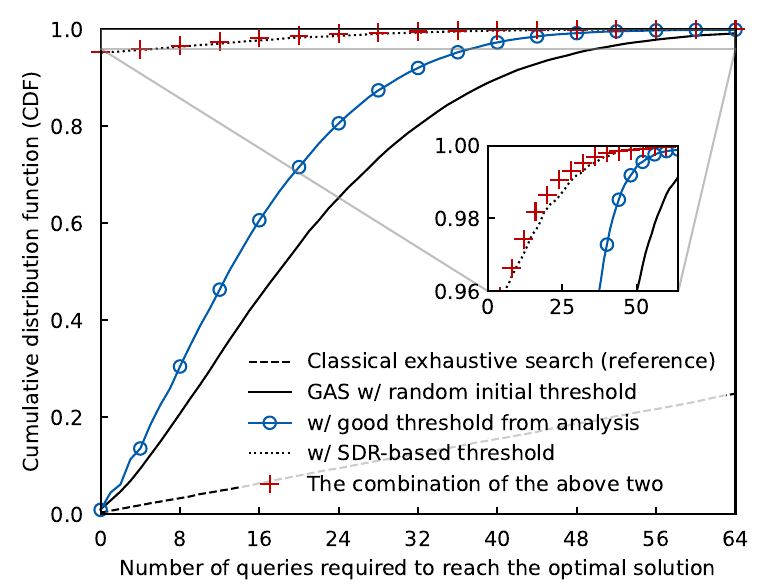}
%\vspace{-3ex}
\caption{Quadratic speedup achieved by GAS in MIMO maximum-likelihood detection. Simply changing the initial threshold inside GAS dramatically alters the query complexity required to reach the optimal solution.}
\label{fig:query}
\end{figure}

\paragraph{Simulation result}
For MIMO detection, Figure~\ref{fig:query} shows the CDF of the number of queries required to reach the optimal solution using different initialization thresholds in GAS.
Compared to classical exhaustive search, which corresponded to optimal detection, the original GAS with a random initial threshold reached the optimum with significantly fewer queries, achieving a quadratic speedup.
Selecting instead good initial thresholds obtained from the distribution analysis or semidefinite relaxation yielded tangible constant-factor reduction in query complexity, and the combination of the two delivered the best performance.

%\vspace{-2ex}

\paragraph{Lesson learned}
Classical results, such as an upper bound on the optimal cost and an estimate (or distribution) of the number of solutions with cost below a given threshold, can further speedup the quantum algorithm with a constant factor. Even wireless researchers can exploit statistical-analysis insights for quantum algorithms.

\subsection{Case Study 2: Combinatorial Optimization in Communication Systems}\label{subsec:case2}

\paragraph{Problem and challenges}

Many discrete optimization problems in communications systems can be mapped to well-known combinatorial optimization such as the graph coloring, quadratic assignment, and traveling salesman problems.
However, certain constraints unique to communication systems make these problems even more complex.
In particular, widely used cost-function formulations are sometimes sub-optimal and can unnecessarily enlarge the search space.
In addition, problems involving max-min (or min-max) objectives cannot be directly expressed as QUBO or HUBO as they are especially difficult to cast as binary polynomials, although such formulations frequently arise in communication systems.

\paragraph{Key contributions}

In \cite{sano2024accelerating}, the graph coloring problem was solved by GAS, by introducing a Gray-code binary encoding to reduce circuit depth leading to the minimization of Hamming distance transitions and the cancellation of redundant quantum gates.
For max-min problems, a QUBO formulation technique was developed exclusively for the max-min dispersion and codebook design problem \cite{yukiyoshi2024quantum}.
Specifically, a power-mean-like approximation replaced the max-min expression, enabling a binary formulation compatible for QC.
In addition, Dicke-state initialization ensured a uniform superposition over feasible solutions without the need for penalty terms, significantly simplifying quantum circuits.

\paragraph{Lesson learned}
Quantum circuits can be simplified significantly through the design intuition of communication engineering.

\section{Conclusion and Discussion
}\label{sec:discussion}

In this article, we introduced QC basics relevant to wireless researchers, highlighting superposition, entanglement, and the state-vector formalism, as well as their synergy with wireless communication theory.
We then presented GAS as an algorithm offering quadratic speedup for binary polynomial optimization.
Our survey was distilled into the common trends, and the two case studies illustrated opportunities for quantum-accelerated communication systems, including synergy with classical computing, pure speedup via appropriate uniform superposition, and circuit simplification.

Several open challenges remain for future exploration at the intersection of quantum and wireless.

\subsection{Exploiting the Harmony Between Quantum and Wireless}
One promising example lies in the Grassmannian packing problem.
In wireless communications, solutions to this problem have been mainly studied in the context of MIMO precoding as well as Grassmannian constellations in noncoherent communications.
These techniques can also be applied to quantum state tomography, where the goal is to reconstruct an unknown quantum state using as few measurements as possible.
Choosing optimal measurement directions that are uniformly spaced is essentially the Grassmannian line packing problem.
However, although several beautifully symmetric measurement directions are known, explicit formulas exist only for limited system sizes.
For larger systems, no general proof has been given, and whether perfect constellations exist at all is still an open problem.

\subsection{Quantum Error Correction}

The fundamental limitation of the Grover-based approach is that FTQC is required for large-scale instances, where idealized, fully error-corrected logical qubits can execute deep circuits with negligible decoherence and gate infidelity.
A key requirement for realizing FTQC is the development of efficient quantum error-correcting codes, prompting interest in structural connections between classical and quantum error correction.
In particular, the relationship between classical LDPC or polar codes and their quantum counterparts is an active area of research.
Such classical codes are widely used in wireless systems for their near-capacity performance and low decoding complexity.
Fundamental differences, such as quantum-specific errors, the no-cloning theorem, and limited connectivity between qubits, make quantum error correction far more challenging.

\subsection{Cost Effectiveness}
A decisive hurdle for the real-world roll-out of quantum technologies is cost-effectiveness.
What ultimately matters to mobile operators is the price per unit time, or per solved instance, which in turn is driven by energy consumption, fabrication cost, and long-term maintenance expenditure.
Ideally, performance should be judged not only by query complexity but by wall-clock time under full system overheads.
These overheads include the latency of data transfer, quantum embedding, read-out delays, and the large qubit overhead imposed by quantum error correction itself.
Only by accounting for this complete cost stack can we decide when FTQC will outperform, in dollars and joules, the best classical alternative.

\section*{Acknowledgment}
This work was supported in part by the Japan Society for the Promotion of Science KAKENHI (JP23K22755 and JP23KK0256), and in part by the Danish National Research Foundation, through the Center of Excellence CLASSIQUE (nr.DNRF187).

\footnotesize{
\bibliographystyle{IEEEtranURLandMonthDiactivated}
\bibliography{main}

@article{fujiwara2024quantum,
  title = {Quantum Speedup for Polar Maximum Likelihood Decoding},
  author = {Fujiwara, Shintaro and Ishikawa, Naoki},
  year = {2024},
  month = nov,
  journal = {arXiv:2411.04727 [quant-ph]},
  doi = {10.48550/arXiv.2411.04727}
}

@article{gilliam2021grover,
  title = {Grover Adaptive Search for Constrained Polynomial Binary Optimization},
  author = {Gilliam, Austin and Woerner, Stefan and Gonciulea, Constantin},
  year = {2021},
  month = apr,
  journal = {Quantum},
  volume = {5},
  publisher = {Verein zur F{\"o}rderung des Open Access Publizierens in den Quantenwissenschaften},
  doi = {10.22331/q-2021-04-08-428}
}

@inproceedings{grover1996fast,
  title = {A Fast Quantum Mechanical Algorithm for Database Search},
  booktitle = {{{ACM}} Symp. {{Theory}} {{Comput.}}},
  author = {Grover, Lov K.},
  year = {1996},
  address = {Philadelphia, Pennsylvania, USA, May 22-24},
  doi = {10.1007/978-1-4419-9863-7_100325},
  isbn = {0-89791-785-5}
}

@article{love2003grassmannian,
  title = {Grassmannian Beamforming for Multiple-Input Multiple-Output Wireless Systems},
  author = {Love, D.J. and Heath, R.W. and Strohmer, T.},
  year = {2003},
  month = oct,
  journal = {IEEE Trans. Inf. Theory},
  volume = {49},
  number = {10},
  pages = {2735--2747},
  doi = {10.1109/TIT.2003.817466}
}

@article{norimoto2023quantum,
  title = {Quantum Algorithm for Higher-Order Unconstrained Binary Optimization and {{MIMO}} Maximum Likelihood Detection},
  author = {Norimoto, Masaya and Mori, Ryuhei and Ishikawa, Naoki},
  year = {2023},
  month = apr,
  journal = {IEEE Trans. Commun.},
  volume = {71},
  number = {4},
  pages = {1926--1939},
  doi = {10.1109/TCOMM.2023.3244924}
}

@article{sano2024accelerating,
  title = {Accelerating {{Grover}} Adaptive Search: {{Qubit}} and Gate Count Reduction Strategies with Higher-Order Formulations},
  author = {Sano, Yuki and Mitarai, Kosuke and Yamamoto, Naoki and others},
  year = {2024},
  journal = {IEEE Trans. Quantum Eng.},
  volume = {5},
  pages = {3101712}
}

@article{yukiyoshi2024quantum,
  title = {Quantum Speedup of the Dispersion and Codebook Design Problems},
  author = {Yukiyoshi, Kein and Mikuriya, Taku and Rou, Hyeon Seok and others},
  journal = {IEEE Trans. Quantum Eng.},
  year = {2024},
  month = aug,
  volume = {5},
  pages = {3103216}
}

@article{hamerly2019experimental,
  title = {Experimental Investigation of Performance Differences between Coherent {{Ising}} Machines and a Quantum Annealer},
  author = {Hamerly, Ryan and Inagaki, Takahiro and McMahon, Peter L. and others},
  year = {2019},
  month = may,
  journal = {Sci. Adv.},
  volume = {5},
  number = {5},
  publisher = {American Association for the Advancement of Science},
  doi = {10.1126/sciadv.aau0823}
}

@article{botsinis2019quantum,
  title = {Quantum Search Algorithms for Wireless Communications},
  author = {Botsinis, Panagiotis and Alanis, Dimitrios and Babar, Zunaira and others},
  year = {Secondquarter 2019},
  journal = {IEEE Commun. Surv. Tutor.},
  volume = {21},
  number = {2},
  pages = {1209--1242},
  doi = {10.1109/COMST.2018.2882385}
}

@article{cuvelier2021quantum,
  title = {Quantum Codes in Classical Communication: {{A}} Space-Time Block Code from Quantum Error Correction},
  author = {Cuvelier, Travis C. and Lanham, S. Andrew and Cour, Brian R. La and others},
  year = {2021},
  journal = {IEEE Open J. Commun. Soc.},
  volume = {2},
  pages = {2383--2412},
  doi = {10.1109/OJCOMS.2021.3121183}
}

@inproceedings{kim2019leveraging,
  title = {Leveraging Quantum Annealing for Large {{MIMO}} Processing in Centralized Radio Access Networks},
  booktitle = {Proc. {{ACM SIGCOMM}}, {{NY}}, {{USA}}},
  author = {Kim, Minsung and Venturelli, Davide and Jamieson, Kyle},
  year = {2019},
  month = aug,
  pages = {241--255},
  doi = {10.1145/3341302.3342072},
  isbn = {978-1-4503-5956-6}
}

@article{gidney2025how,
  title = {How to Factor 2048 Bit {{RSA}} Integers with Less than a Million Noisy Qubits},
  author = {Gidney, Craig},
  year = {2025},
  month = may,
  journal = {arXiv:2505.15917},
  doi = {10.48550/arXiv.2505.15917}
}

@article{bian2015differential,
  title = {Differential Spatial Modulation},
  author = {Bian, Yuyang and Cheng, Xiang and Wen, Miaowen and Yang, Liuqing and Poor, H. Vincent and Jiao, Bingli},
  year = {2015},
  journal = {IEEE Trans. Veh. Technol.},
  volume = {64},
  number = {7},
  pages = {3262--3268},
  doi = {d},
  isbn = {0018-9545 VO - 64}
}

@article{miyake2001geometric,
  title = {Geometric Strategy for the Optimal Quantum Search},
  author = {Miyake, Akimasa and Wadati, Miki},
  year = {2001},
  month = sep,
  journal = {Phys. Rev. A},
  volume = {64},
  number = {4},
  pages = {042317},
  publisher = {American Physical Society},
  doi = {10.1103/PhysRevA.64.042317}
}

@article{tang2022dequantizing,
  title = {Dequantizing Algorithms to Understand Quantum Advantage in Machine Learning},
  author = {Tang, Ewin},
  year = {2022},
  month = nov,
  journal = {Nat. Rev. Phys.},
  volume = {4},
  number = {11},
  pages = {692--693},
  publisher = {Nature Publishing Group},
  doi = {10.1038/s42254-022-00511-w},
  copyright = {2022 Springer Nature Limited}
}
}

\vspace{-1cm}
\begin{IEEEbiographynophoto}{Naoki Ishikawa} [SM'22] is an Associate Professor at Yokohama National University, Japan. He is an Associate Editor of IEEE TVT.
\end{IEEEbiographynophoto}
\vspace{-1cm}
\begin{IEEEbiographynophoto}{Giuseppe T. F. de Abreu} [SM'09] is a Professor at Constructor University Bremen, Germany. He is an Editor of IEEE SPL and IEEE OJ-COMS.
\end{IEEEbiographynophoto}
\vspace{-1cm}
\begin{IEEEbiographynophoto}{Petar Popovski} [F'16] is a Professor at Aalborg University, Denmark and a Visiting Excellence Chair at the University of Bremen. He is the Editor-in-Chief of IEEE JSAC.
\end{IEEEbiographynophoto}
\vspace{-1cm}
\begin{IEEEbiographynophoto}{Robert W. Heath, Jr.} [F'11] is a Professor at University of California, San Diego, CA, USA. He is also the President and the CEO of MIMO Wireless Inc. He is a member of the US National Academy of Engineering.
\end{IEEEbiographynophoto}

\end{document}